\documentclass[12pt]{article} 

\begin{document} 
\topmargin 0pt 
\oddsidemargin 0mm
\renewcommand{\thefootnote}{\fnsymbol{footnote}}
\begin{titlepage}
\vspace{5mm}

\begin{center}
{\Large \bf The contribution to the perihelion advance of Mercury
  arising out of the dependence of mass on gravitational potential } 
\vspace{6mm}

{\large Harihar Behera{\footnote{email: harihar@iopb.res.in}}}\\
\vspace{5mm}
{\em
Department of Physics, Utkal University, Vanivihar,
Bhubaneswer-751004,Orissa, India\\}
\vspace{3mm}
\end{center}
\vspace{5mm}
\centerline{{\bf {Abstract}}}
\vspace{5mm}
The effect of general relativistic prediction of the dependence of
mass on gravitational potential on the dynamics of a planet moving
around the sun is shown to have a negative contribution of $
14\cdot\,326\,\mbox{arcsec/century}\,$ towards the overall
non-Newtonian perihelion advance of Mercury. \\


\end{titlepage}
\section{Introduction}
The general relativistic effect of the dependence of mass on
gravitational potential \cite{1,2} taken in conjunction with the Schwarzschild
metric has got the virtue of addressing \cite{1} from a single point
of view four effects,viz.,\\
(a) the gravitational red-shift of photons flying away from a
gravitating body,\\
(b)the increase of energy difference between levels in atoms with the
increase of their distance from the gravitating body,\\
(c) the retardation of radar echo from planets,\\
(d) the deflection of star light observed during a solar eclipse,\\
which are usually discussed separately in the literature. But if one
wishes to see the effect of the dependence of
mass on gravitational potential on the dynamics of a planet moving
around the sun, then a negative contribution of $
14\cdot\,326\,\mbox{arcsec/century}\,$ towards the overall
non-Newtonian perihelion advance of Mercury seems inevitable which deserves attention in the standard general
relativistic explanation for the observed anomalous perihelion advance 
of Mercury.\\

\section{Perihelion Advance from Schwarzschild Solution }   
The standard general relativistic explanation for the observed
non-Newtonian perihelion advance of a planet's orbit may be obtained from the
so-called Schwarzschild spherically symmetric solution of Einstein's
field equations corresponding to an additional Hamiltonian in the
bound Kepler problem \cite{3} of the form :
\begin{equation}
\bigtriangleup H\,=\,-\,\frac{h_{E}}{r^{3}}
\end{equation} 
with 
\begin{equation}h_{E}\,=\,\frac{kl^{2}}{m^{2}c^{2}},\,\,\,\,\,\,\,\,\,\,\,\,{\,k\,=\,GM_{\odot}m \,}\,
\end{equation}
where $\,l\,,\,m\,$ represent the angular momentum and mass of the
planet respectively,$\,M_{\odot}\,$ is the solar mass,$\,G\,$ is
Newton's Universal gravitational constant; $\,r\,$ is the distance
between the Sun and the planet and $\,c\,$ is the speed of light in
vacuum.\\
Thus the potential energy after the correction $(1)$ is
\begin{equation} U\,=\,-\,\frac{k}{r}\,-\,\frac{h_{E}}{r^{3}}
\end{equation}
It is known \cite{3} that if  a potential with
$\,1/r^{3}\,$  form
is added to a central force perturbation of the bound Kepler problem,
the orbit in the bound problem is an ellipse in a rotating coordinate
system. In effect the ellipse rotates, and the periapsis appears to
precess.If the perturbation Hamiltonian is $(1)$,then it predicts \cite{3} a precession of the perihelion of a planet at an
average  rate of
\begin{equation} 
\dot{\tilde{\bf\omega}}\,=\,\frac{6 \pi\,k\,m^{2}h_{E}}{\,\tau\,l^{4}}
\end{equation}
where $\,\tau\,$ is the classical period of revolution of the planet
around the sun.With
$\,l^{2}\,=\,GM_{\odot}\,m^{2}\,a\,(\,1\,-\,e^{2})\,$ and
$\,h_{E}\,=\,\frac{GM_{\odot}l^{2}}{mc^{2}}\,$,the expression $(4)$
represents Einstein's expression for the non-Newtonian perihelion
advance of a planet's orbit \cite{3}:
\begin{equation}
\dot{\tilde{\bf\omega}}_{E}\,=\,\frac{6\,\pi\,k^{2}\,}{\,\tau\,l^{2}\,c^{2}}\,=\, \frac{6\,\pi\,GM_{\odot}}{\,\tau\,c^{2}\,a\,(\,1\,-\,e^{2}\,)}
\end{equation}
where the symbols have their usual meanings.\\
\section{Contribution to the perihelion advance of Mercury
  arising out of the dependence of mass on gravitational potential }
One of the predictions of general relativity is that the rest energy
$\,E_{0}\,$ of any massive object increases with increase of the
distance from the gravitating body because of the increase of the
potential $\phi$ \cite{1,2}:
\begin{equation}E_{0}\,=\,mc^{2}\left(\,1\,+\,\frac{\phi}{c^{2}}\right)
\end{equation}
The gravitational potential is
\begin{equation}\phi\,=\,-\frac{GM_{\odot}}{r}
\end{equation}
where the symbols have their usual meanings. Because of the relations
$\,E\,=\,mc^{2}\,$ and  $(6)-(7)$, the gravitationally
modified mass of a planet in the gravitational field of the sun may be 
expressed as
\begin{equation}m^{\star}\,=\,m\left(\,1\,-\,\frac{GM_{\odot}}{c^{2}r}\right)
\end{equation}
If we wish to see the effect of $(8)$ on the dynamics of a planet
moving around the sun, then instead of $(3)$ we should consider an
equation of the form :
\begin{equation} U^{\star}\,=\,-\,\frac{k^{\star}}{r}\,-\,\frac{h^{\star}_{E}}{r^{3}}
\end{equation}
where
\begin{equation} \,k^{\star}\,=\,GM_{\odot}m^{\star}\,=\,k\left(\,1\,-\,\frac{GM_{\odot}}{c^{2}r}\right),\,\,\,\,
and \,\,\,h^{\star}_{E}\,=\,h_{E}
\end{equation}
since $\,h_{E}\,=\,\frac{(GM_{\odot})^{2}a(1-e^{2})}{c^{2}}$ is
independent of the mass of the planet. Therefore  $(9)$ may be
rewritten as
\begin{equation} U^{\star}\,=\,-\,\frac{k}{r}\,-\,\frac{h}{r^{2}}-\,\frac{h_{E}}{r^{3}}
\end{equation}
where 
\begin{equation} h\,=\,-\,\frac{kGM_{\odot}}{c^{2}}
\end{equation}
It is now apparent from $(11)$ that the consideration of the
dependence of mass on gravitational potential introduced an additional 
Hamiltonian of the form $\,\frac{1}{r^{2}}\,$ into the equation $(3)$. In
consequence of this effect, a new contribution towards the overall
non-Newtonian perihelion advance of a planet's orbit may be estimated
\cite{3} at
\begin{equation} 
\dot{\tilde{\bf\omega}}_{new}\,=\,\frac{2\pi\,m\,h}{\,\tau\,l^{2}}\,=\,-\,\frac{2\,\pi\,GM_{\odot}}{\,\tau\,c^{2}\,a\,(\,1\,-\,e^{2}\,)}\,=\,-\,\frac{\dot{\tilde{\bf\omega}}_{E}}{3}
\end{equation}
For Mercury, the value of
$\,\dot{\tilde{\bf\omega}}_{E}\,=\,42\cdot98\,$ arcsec/century - a
well known data \cite{4,5,6}. So
$\dot{\tilde{\bf\omega}}_{new}\,=\,-\,14\cdot326\,$ arcsec/century,
for Mercury. This is in no way a negligible contribution.\\

\section{Discussion}

The Modern observational value of the anomalous perihelion advance of
Mercury is at $ \dot{\tilde{\bf\omega}}\,\approx\,43'' $ per
century \cite{6}. For Mercury, General Relativity predicts this
phenomenon at                                                                  \begin{equation} 
\dot{\tilde{\bf\omega}}_{GR}\,=\,\left[
  42\cdot98''\,+\,1\cdot289''(\,J_{2\odot}/{10}^{-5}\,)\right]\,\,{\rm{per\,\,century},}
\end{equation}
where $\,J_{2\odot}$ is the magnitude of solar quadrupole moment of the 
sun - a definite  value of which has still to be determined \cite{4}. Although
the Solar
quadrupole moment contribution is a Newtonian contribution, it was
taken account of because it was not there in the standard
expression. It is to be noted that the standard explanation for the
perihelion advance of Mercury offered by $(14)$ does not include the
contribution arising out of the dependence of mass on gravitational
potential discussed here. But if one takes this effect into account ,
then $(14)$
taken in conjunction with $(13)$ is not in agreement with the observed
perihelion advance of Mercury. However , as a remedy for this disagreement one may suggest a  value of  $ J_{2\odot} $ at
\begin{equation} J_{2\odot}\,\approx\,1\cdot11\times{10}^{-4} 
\end{equation}
Unfortunately, such a suggestion is not in agreement with none of the
existing observational data on $ J_{2\odot} $. For example, Dicke and
Goldenberg's \cite{7} optical measurements of the solar oblateness
showed
$J_{2\odot}\,\simeq\,(2\cdot5\,\pm\,0\cdot2)\times{10}^{-5}\,$. It is to be noted that
Dicke and Goldenberg's measured value of $ J_{2\odot} $ makes the
observations of the perihelion advance of Mercury disagree with the
predictions of General Relativity \cite{4}. In 1974, Hill
et.al., \cite{8} , by measuring Sun's optical oblateness, obtained $
J_{2\odot}\,\simeq\,(\,0\cdot1\,\pm\,0\cdot4\,)\times\,{10}^{-5}\,$. The 
values of $ J_{2\odot} $ much smaller than the above cited values have 
also been measured by others ( for
references see for example \cite{4}). From these considerations, the suggested  value of $
J_{2\odot} $ in $(15)$ is unacceptable. In this scenario, the the
effect of the dependence of mass on gravitational potential seems 
to make a point of concern for the standard general relativistic 
explanation for the observed anomalous perihelion advance of Mercury
in particular and other planets in general which merits further study.\\

\textbf{Acknowledgments}\\ 
  
 The author is indebted to Prof. N. Barik and Prof. L. P. Singh of
 Utkal University, Vanivihar, Bhubaneswer for fruitful discussion and
 valuable suggestions.The author also acknowledges the help received from Institute of Physics,Bhubaneswar for using its Library and Computer Centre for this
 work.\\ 
 
\end{document}